# Parallel Bayesian Additive Regression Trees

Matthew T. Pratola

Statistical Sciences Group, Los Alamos National Laboratory

and

Hugh A. Chipman

Department of Mathematics and Statistics, Acadia University

and

James R. Gattiker

Statistical Sciences Group, Los Alamos National Laboratory

and

David M. Higdon*

Statistical Sciences Group, Los Alamos National Laboratory

and

Robert McCulloch

Booth School of Business, University of Chicago

and

William N. Rust

Statistical Sciences Group, Los Alamos National Laboratory

September 1, 2013

## Abstract

Bayesian Additive Regression Trees (BART) is a Bayesian approach to flexible non-linear regression which has been shown to be competitive with the best modern predictive methods such as those based on bagging and boosting. BART offers some advantages. For example, the stochastic search Markov Chain Monte Carlo (MCMC) algorithm can provide a more complete search of the model space and variation across MCMC draws can capture the level of uncertainty in the usual Bayesian way. The BART prior is robust in that reasonable results are typically obtained with a default prior specification. However, the publicly available implementation of the BART algorithm in the R package BayesTree is not fast enough to be considered interactive with over a thousand observations, and is unlikely to even run with 50,000 to 100,000 observations. In this paper we show how the BART algorithm may be modified and then computed using single program, multiple data (SPMD) parallel computation implemented using the Message Passing Interface (MPI) library. The approach scales nearly linearly in the number of processor cores, enabling the practitioner to perform statistical inference on massive datasets. Our approach can also handle datasets too massive to fit on any single data repository.

*This work was supported in part by the U.S. Department of Energy Office of Science, Office of Advanced Scientific Computing Research, Scientific Discovery through Advanced Computing (SciDAC) program.

# 1 Introduction

The challenges confronting modern statistics are often very different from those faced by classical statistics. For instance, in today's applied problems one often has huge datasets involving millions of observations and hundreds or thousands of variables. Examples include the high-dimensional simulators found in computer experiments (Higdon et al., 2008; Pratola et al., 2013), large complex genetic datasets in biomedicine (Sinha et al., 2009) and massive datasets of consumer behavior in computational advertising (Agarwal et al., 2010). In such problems, one challenge is modeling complex data structures in a manner that is both efficient and lends itself to useful inference.

The Bayesian Additive Regression Tree (BART) model is a Bayesian Non-Parametric model that was presented in Chipman, George, and McCulloch (2010) (henceforth CGM). CGM consider the fundamental model

$$Y = f(x) + \epsilon, \qquad \epsilon \sim N(0, \sigma^2)$$

where $x = (x_1, \ldots, x_d)$ represents $d$ predictors. The function $f$ is represented as a sum of regression tree models $f(x) = \sum_{j=1}^{m} g_j(x)$ where $g_j(x)$ represents the contribution to the overall fit provided by the $j^{th}$ regression tree. The number of regression tree models, $m$, is chosen to be large and the prior constrains the contribution of each regression tree model so that the overall fit is the sum of many small contributions. A Markov Chain Monte Carlo algorithm draws from the full joint posterior of all the regression tree models and $\sigma$.

CGM compare the out-of-sample predictive performance of BART with a boosting method, Random Forests, support vector machines, and Neural Nets and report that BART's performance is competitive. However, the data sets used for comparison have sample sizes of at most $n \approx 15,000$ (the drug-discovery example). All the BART computations reported in CGM are done using the function `bart` in the R (R Core Team, 2012) package `BayesTree`. For the larger sample sizes often confronted in modern statistical applications, the `bart`/`BayesTree` implementation is hopelessly slow and uses large amounts of memory.

In this paper we describe an implementation of the BART method which is able to handle much larger data sets and is much faster for moderately sized data sets. Our approach, available at `http://www.rob-mcculloch.org`, first simplifies the C++ representation of the regression tree models. We then simplify the Metropolis Hastings step in the BART MCMC algorithm presented in CGM. Finally, and crucially, we show that with our simplified BART, a Single Program, Multiple Data (SPMD) parallel implementation can be used to dramatically speed the computation. Our lean model representation is propagated across all processor cores while the data is split up into equal portions each of which is allocated to a particular core. Computations on the entire data set are done by computing results for each data portion on its own core, and then combining results across portions. The approach scales nearly linearly in the number of processor cores, and can also handle datasets too massive to fit on any single data repository.

While prediction given a set of out-of-sample $x$'s is fairly straightforward using BART, interpreting the fit may be difficult. Methods for interpretation often rely on designing a large number of $x$ at which to make predictions. We also show how these prediction computations may be done with SPMD parallel computation.

The rest of the paper proceeds as follows. Section 2 reviews enough details of the MCMC algorithm for fitting BART so that the reader may understand how the three simplifications described above are carried out. Section 3 explains how we have implemented an efficient and parallel version of the BART algorithm. Section 4 compares the actual times needed to run the BART implementation of this paper done serially, and the parallel BART implementation. Section 5 details the scalability of our parallel BART implementation. Section 6 discusses parallel computation for large numbers of predictions. Section 7 presents an example of analyzing data sets with large sample sizes using BART. Finally, we summarize our findings in Section 8

## 2 The BART Algorithm

We briefly review those aspects of the BART methodology in CGM necessary for the understanding of this paper. We borrow liberally from CGM.

## 2.1 The Sum of Trees Model

The BART model expresses the overall effect of the regressors as the sum of the contributions of a large number of regression trees. In this section we provide additional notation and detail so that we can describe our approach.

We parametrize a single regression tree model by the pair $(T, M)$. $T$ consists of the tree nodes and decision rules. The splits in $T$ are binary so that each node is either terminal (at the bottom of the tree) or has a left and right child. Associated with each non-terminal node in $T$ is a binary decision rule which determines whether an $x$ descends the tree to the left or to the right. Typically, decision rules are of the form "go left if $x_v < c$" where $x_v$ is the $v$th component of $x$. Let $b = |M|$ be the number of terminal nodes. Associated with the $k^{th}$ terminal node is a number $\mu_k$. $M$ is the set of terminal node $\mu_k$ values: $M = (\mu_1, \ldots, \mu_b)$. To evaluate a single regression tree function, we drop an $x$ down the tree $T$ until it hits terminal node $k$. We then return the value $\mu_k$. We use the function $g(x; T, M)$ to denote this returned value $\mu_k$ for input $x$ and tree $(T, M)$.

Let $(T_j, M_j)$ denote the $j^{th}$ regression tree model. Thus the $g_j(x)$ introduced in Section 1 is expressed as $g_j(x) \equiv g(x; T_j, M_j)$. Our sum of trees model is

$$Y = \sum_{j=1}^{m} g(x; T_j, M_j) + \epsilon, \qquad \epsilon \sim N(0, \sigma^2). \tag{1}$$

The prior specification for the parameter $((T_1, M_1), \ldots, (T_m, M_m), \sigma)$ is key to the BART methodology. Since our focus here is on computation, the reader is referred to CGM for the details.

## 2.2 The BART MCMC Algorithm

At the top level, the BART MCMC is a simple Gibbs sampler. Let $T_{(j)}$ denote all the trees except the $j^{th}$ and define $M_{(j)}$ similarly. Our Gibbs sampler then consists of iterating the draws:

$$(T_j, M_j) \,|\, T_{(j)}, M_{(j)}, \sigma, y, \qquad j = 1, 2, \ldots, m \tag{2}$$

$$\sigma \,|\, T_1, \ldots T_m, M_1, \ldots, M_m, y.$$

The draw of $\sigma$ is straightforward since given all the $(T_j, M_j)$, (1) may be used to calculate $\epsilon$, which can be treated as an observed quantity.

To make each of the $m$ draws of $(T_j, M_j)$, note that the conditional distribution $p(T_j, M_j | T_{(j)}, M_{(j)}, \sigma, y)$ depends on $(T_{(j)}, M_{(j)}, y)$ only through

$$R_j \equiv y - \sum_{k \neq j} g(x; T_k, M_k),$$

the $n-$vector of partial residuals based on a fit that excludes the $j$th tree. Conditionally, we have the single tree model

$$R_j = g(x; T_j, M_j) + \epsilon.$$

Thus, the $m$ draws of $(T_j, M_j)$ given $(T_{(j)}, M_{(j)}, \sigma, y)$ in (2) are equivalent to $m$ draws from

$$(T_j, M_j) \,|\, R_j, \sigma, \qquad j = 1, \ldots, m,$$

and each one of these draws may be done using single tree methods.

Each single tree model draw of $(T_j, M_j)$ is done using the approach of Chipman, George, and McCulloch (1998). The prior specification is chosen so that we can draw from the joint $(T_j, M_j) \,|\, R_j, \sigma$ by first analytically integrating out $M_j$ and drawing from the marginal $T_j \,|\, R_j, \sigma$ and then drawing from the conditional $M_j \,|\, T_j, R_j, \sigma$.

The draws $T_j \,|\, R_j, \sigma$ are the heart of the algorithm. It is in these steps that the structure of the trees change. These draws are carried out using a Metropolis-Hastings step. Given a current tree structure, a modification is proposed and the modification is accepted or rejected (in which case the current structure is retained) according to the usual MH recipe.

In CGM, four different tree modification proposals are used. First, there are a complementary pair of BIRTH/DEATH proposals. A BIRTH proposal picks a terminal node and proposes a decision rule so that the node gives birth to two children. A DEATH step picks a pair of terminal nodes having the same parent node, and proposes eliminating them so that the parent becomes a terminal node.

The CHANGERULE move leaves the parent/child structure of the tree intact, but proposes a modification to the decision rule associated with one of the non-terminal nodes. The SWAP move picks a pair of non-terminal parent/child nodes and proposes swapping their decision rules.

The CGM method randomly picks one of the four tree modification proposals.

# 3 Efficient and Parallel Computation

In this section we recall the historical usage of parallel computing in Bayesian methods, and detail how we have efficiently coded and simplified the BART algorithm to implement a single program, multiple data (SPMD) parallel computation.

## 3.1 Background

The use of parallel computing to implement efficient statistical models has received increasing attention as the real-world datasets requiring statistical modeling and analysis become larger and increasingly complex. In likelihood-based inference, there are many approaches one can apply to speed up computation of the objective function, such as using conditional independence to factor the likelihood into more tractable components, using software libraries for solving linear systems in parallel, and in the case of some Gaussian models, working with the precision matrix which is typically sparse, making for easier computations.

The ability to use such tricks in the context of big data has been aided with the advent of frameworks such as MapReduce (Dean and Ghemawat, 2008), a restricted form of master/slave parallel computation that removes the many details inherent in handling large datasets and writing efficient parallel algorithms. The MapReduce framework has seen success in some applications of machine learning and statistics (Chu et al., 2007). This approach to handling large datasets usually results from relying on i.i.d. components of the dataset (e.g. replicate observations) so that the effective dimensionality is much smaller than the full dataset (e.g. Kleiner et al. (2012)), or fitting the same model to different subsets of the full dataset and then recombining these fits to approximate the full likelihood estimates.

Using parallel computations in a Bayesian modeling framework is more complex. The first challenge is that most practically relevant Bayesian models do not have closed-forms for the posterior distribution, and so usually rely on MCMC-based inference, which is already computationally demanding and not necessarily straightforward to implement in terms of achieving efficient sampling. Some approaches include parallel Markov Chains, blocking and marginalisation, with the latter two often involving a tradeoff between more efficient sampling performance and more efficient computational performance (Wilkinson, 2005).

These strategies for parallelising single chains or multiple chains have been discussed extensively (Wilkinson, 2005; Rosenthal, 2000), however in the context of big data the practical ability to perform Bayesian Inference using such approaches is less clear. A possibility is to implement such algorithms using MapReduce. Yet, due to the iterative nature of MCMC algorithms and the requirement to store intermediate state information in local memory during the computation of the Markov chain, the MapReduce framework is not a feasible approach. This is because MapReduce was developed for making simple data-parallel applications that require simple computations on a *single* pass of a large distributed dataset.

For example, the MapReduce specification (Dean and Ghemawat, 2008) states that outputs of the Map step are written to disks local to the Map workers. Then, the Reduce workers access these intermediate files, which are local to the Map workers, remotely over the network. Next, the Reduce workers sort these intermediate output files. Finally, the sorted intermediate outputs are passed to the users Reduce function, after which the final output is again written to local disk. In an MCMC application, this would be terribly inefficient as each iteration of the MCMC would require all these local disk writes, reads and sorts. As such, MapReduce has not been used for Bayesian MCMC-based inference in a big data setting, although research into more general frameworks that enable iterative computations is ongoing (Ekanayake et al., 2010).

Our particular interest is in scalable and flexible regression methods for large high-dimensional datasets, for which we have found the BART model to be well suited. An additional challenge in making such a Bayesian Non-Parametric model scalable lies in the unbounded dimensionality of such models, which can lead to large amounts of communication overhead (Doshi-Velez et al., 2010). However, as we outline in the next section, the BART parallel sampler we devise uses the Bayesian Non-Parametric form of the model to its advantage, leading to a scalable and efficient sampler.

## 3.2 Parallel, Scalable BART

Our first step was to code the tree models as simply as possible. Each node in a tree is represented as an instance of a C++ class. The class has only six data members: (i) a mean $\mu$ (ii) an integer $v$ and integer $c$ such that the decision rule is left if $x_v < c^{th}$ cutpoint (iii) a pointer to the parent node (iv) pointers to left and right children. Note that for a terminal node the pointers to left and right children are not assigned. For each component of $x$ a discrete set of possible cutpoints are pre-calculated so that a cutpoint may be identified with an integer. This is the minimal information needed to represent a regression tree. A minimal representation speeds computation in that when trees are dynamically grown and shrunk as the MCMC runs, little computation is needed to make the modifications. In addition, in our SPMD implementation tree modifications must be propagated across the cores so that minimizing the amount of information that must be sent is important. A consequence of this lightweight representation is that some quantities that characterize a tree must be recomputed on demand, rather than stored. For example, to determine the depth of a node, pointers to parents, grandparents, etc. must be followed. However, these computations are fast, due in part to the typically small number of nodes in the trees used. Note that the implementation in the `R` function `bart` (package `BayesTree`), C++ classes are also used to represent nodes in a tree. However, the C++ classes in that implementation are much more complicated so that more computation and memory is needed to maintain them.

Our second simplification relative to CGM (and `bart/BayesTree`) is that only the BIRTH and DEATH tree modification proposals are used. In Chipman, George, and McCulloch (1998) it was found that using only BIRTH/DEATH tree modification moves resulted in an inferior MCMC exploration of the model space. Results obtained with different random number generator seeds could be dramatically different. However, these findings were in the context of a single regression tree model. BART behaves in a fundamentally different way, with individual trees that typically contain far fewer nodes. Small trees correspond to a more easily searched space. We have found, in many examples, that the fits obtained using only the BIRTH/DEATH moves are extremely similar to those obtained using all four moves discussed above in Section (2). A similar finding was also seen in a dynamic trees context (Taddy et al., 2011). It is possible to efficiently code additional proposals that are data-independent, and hence easily implemented in our data-parallel framework (Pratola, submitted), however our goal here was to do things as simply as possible.

Even these first two modifications result in a very noticeable improvement in the serial performance of our BART MCMC sampler. For example, in Table 1 the performance improvement for the serial sampler is seen to be from 4 times faster for small sample sizes up to 6 times faster for larger sample sizes. As such, the parallel results subsequently reported in this paper will be compared to the new serial BART sampler.

Table 1: Performance of `bart/BayesTree` serial MCMC sampler versus the new serial MCMC sampler for moderately sized datasets. Both samplers were run on a simulated dataset with 5 covariates using 2,000 MCMC iterations with the first 1,000 discarded as burn-in.

| n | `bart/BayesTree` MCMC | new MCMC |
|---|---|---|
| 1,000 | 57.725 | 14.057 |
| 2,000 | 136.081 | 27.459 |
| 4,000 | 298.712 | 54.454 |
| 6,000 | 463.861 | 82.084 |
| 8,000 | 651.683 | 107.911 |
| 10,000 | 817.711 | 135.764 |

Finally, we outline our SPMD parallel computation. Given $p+1$ processor cores numbered $0, 1, 2, \ldots, p$,

we split the data $(y, x)$ into $p$ (approximately) equally-sized portions, $(y_{(1)}, x_{(1)}), \ldots, (y_{(p)}, x_{(p)})$ where the $i^{th}$ data portion resides on core $i$. The current state of the regression tree models $((T_1, M_1), (T_2, M_2), \ldots, (T_m, M_m))$ is copied across all $p+1$ cores. The algorithm proceeds in a master-slave arrangement, where core 0 contains no observed data and only manages the MCMC sampler, while all computations involving the observed data take place on the $p$ slave cores in parallel. Figure 1 illustrates the setup. Each large rectangle in the figure represents a core. Within each core, multiple trees are depicted representing the $(T_j, M_j)$. However, core $i$ only has data portion $y_{(i)}, x_{(i)}$.

As a simple example consider the draw $\sigma \mid T_1, \ldots T_m, M_1, \ldots, M_m, y$. To make this draw we just need the sufficient statistic $\sum_{i=1}^{n} \epsilon_i^2$, where $n$ is the total number of observations and $\epsilon_i = y_i - \sum_{j=1}^{m} g(x_i; T_j, M_j)$. Since each core has copies of all the $(T_j, M_j)$ it can compute the $\epsilon_i$ for its data portion and sum their squares. To make the draw of $\sigma$, the master node sends a request out to each slave core and each core responds with its portion of the total residual sum of squares. The master core adds up the residual sums of squares portion received from each slave and then draws $\sigma$. The ability to decompose sufficient statistics into sums of terms corresponding to different parts of the data enables this SPMD approach.

Consider the case of a BIRTH step. A particular terminal node of tree $T_j$ in our sum of trees model has been chosen. A candidate decision rule (given by a choice of $(v, c)$) has been proposed. If we accept the move, the terminal node will be assigned the decision rule, and will be given a left and right child (and will hence cease to be a terminal node). To evaluate the MH acceptance probability of this proposed tree modification we need only know the sum of the partial residuals $R_j$ for the observations assigned to the new left child and the sum for the observations in the new right child. This simplification is again the result of sufficiency under the assumption of normal errors. The master node manages the MH step. To compute the partial residual sums, the master sends out requests to the slaves and then sums the partial sums of the partial residuals. If the move is accepted, the master node then must propagate the change in $T_j$ and $M_j$ out to all the slaves.

The overall parallel MCMC sampler is summarized in Table 2. The calculations and communications required for each step of the MCMC are summarized by describing each operation performed on the Master node and on a given Slave node. The number of bytes for communication operations are specified as s(# bytes) and r(# bytes) for sends and receives respectively. Note that at each MCMC iteration, the BIRTH/DEATH proposals for $T_j \mid R_j, \sigma$ draws will involve at most $m$ tree modifications that must be propagated across the slaves, depending on how many MH proposals are accepted. The sufficient statistics needed for the left/right nodes to undergo BIRTH/DEATH are denoted with subscripts $l, r$ in the table.

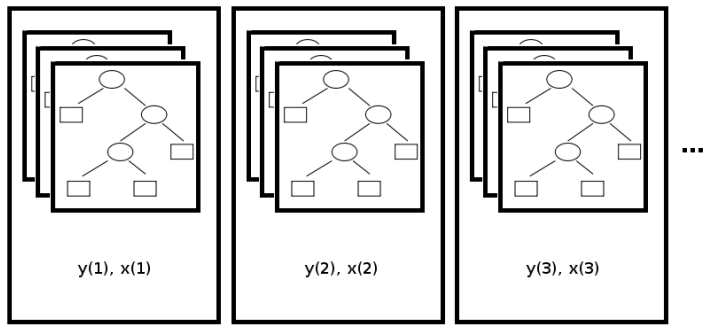

Figure 1: Each core has only a portion of the complete dataset $(y, x)$ but all the model fit information (represented here by the trees).

The $M_j \mid T_j, R_j, \sigma$ draws will involve propagating $\sum_{j=1}^{m} b_j$ new $\mu$ values across the slaves. The current value of $\sigma$ need only be maintained on the master core, so the communication overhead in this step comes from receiving the partial residual sums-of-squares (RSS) from the slaves.

Note that none of the parallel communications outlined depend on the sample size of the data set, and all but two are a small constant number of bytes. Conditional on the tree model being accessible on each core, all expensive computations involving the actual data (e.g. calculation of the partial sufficient statistics) are performed independently on each slave core operating solely on the subset of data assigned to that core. Because of our lean model representation this algorithm is able to sample from the posterior distribution $p((T_1, M_1), \ldots, (T_m, M_m), \sigma | y)$ efficiently with little communication overhead between computing cores.

The notions of sufficiency and the reduction of data to a statistical model figure prominently in this efficient implementation of BART. The large volumes of data are characterized by a few sufficient statistics and the simple statistical model, giving a compressed representation of the data that can be held in each cores local memory. This enables the quick exploration of the model space with the parallel BART algorithm.

## 4 Timing Results

Here we look at how this MCMC implementation speeds up with additional processors, considering a single dataset $(x, y)$ where $y$ is a 200,000-vector and $x$ is 200,000×40. These data are produced by a realization of the random function generator of Friedman (2001). The entries of $x$ are i.i.d. draws from a $U[-1, 1]$ distribution. Briefly, given $x_i$, a row of $x$, $y_i$ is an additive combination of randomly produced normal kernels

$$y_i = \sum_{\ell=1}^{q} a_\ell q_\ell(x_i) + \epsilon_i. \tag{3}$$

| Op (bytes) | Master | Slave | Op (bytes) |
|---|---|---|---|
| \multicolumn{4}{c}{$T_j\|R_j,\sigma \quad \forall j=1,\ldots,m$} ||||
| | BIRTH | BIRTH | |
| s(12) | Proposed split node, variable and cutpoint | Split node, variable and cutpoint | r(12) |
| | | Calculate partial suff. stat. | |
| r(24) | Suff. stat. | $n_l, n_r, \sum R_l, \sum R_r$ | s(24) |
| | MH Step | | |
| s(28) | If accept BIRTH: node, variable, cutpoint, $\mu_l, \mu_r$ | Update node, variable, cutpoint, $\mu_l, \mu_r$ | r(28) |
| s(0) | Else reject BIRTH signal | Else reject BIRTH signal | r(0) |
| | DEATH | DEATH | |
| s(8) | Nodes of children to kill | Nodes of children to kill | r(8) |
| | | Calculate partial suff. stat. | |
| r(24) | Suff. stat. | $n_l, n_r, \sum R_l, \sum R_r$ | s(24) |
| | MH Step | | |
| s(28) | If accept DEATH: new terminal node and $\mu$ | Update new terminal node and $\mu$ | r(28) |
| s(0) | Else reject DEATH signal | Else reject DEATH signal | r(0) |
| \multicolumn{4}{c}{$M_j\|T_j, R_j, \sigma \quad \forall j=1,\ldots,m$} ||||
| | | Calculate partial suff. stat. for all $b_j$ bottom nodes | |
| r($20b_j$) | Suff. stat. | $\{n_i, R_i, R_i^2\}_{i=1}^{b_j}$ | s($20b_j$) |
| | Gibbs Step | | |
| s($8b_j$) | Gibbs draw of $M_j$ | Update $M_j$ | r($8b_j$) |
| \multicolumn{4}{c}{$\sigma\|\cdot$} ||||
| | | Calculate partial RSS | |
| r(8) | RSS | $\sum \epsilon^2$ | s(8) |
| | Gibbs draw of $\sigma$ | | |

Table 2: Summary of parallel MCMC sampler. The left "Op" column summarizes the communication operations ({s}end/{r}eceive) and number of bytes for the master node while the right "Op" column summarizes operations for the slave nodes.

The coefficients $a_\ell$ are i.i.d. $U[-1, 1]$ draws. We take $q = 30$ and $\epsilon_i$ to be i.i.d. $N(0, \sigma^2)$ with $\sigma = 0.15$. The normal kernels $q_\ell(x)$ are determined by first randomly selecting a subset of components $[\ell]$ of $x$, giving $x_{[\ell]}$, randomly rotating these component directions with rotation matrix $U_\ell$, and then stretching or dilating these rotated components according to the diagonal matrix $D_\ell$

$$q_\ell(x) = \exp\left\{-\frac{1}{2}(x_{[\ell]} - \mu_\ell)^T U_\ell D_\ell^{-1} U_\ell^T (x_{[\ell]} - \mu_\ell)\right\}.$$

The mean vectors $\mu_\ell$ are independent $U[-1, 1]$ draws, same as the $x_{[\ell]}$'s. The diagonal matrix $D_\ell$ has diagonal entries $d_k$, with $\sqrt{d_k} \sim U[.1, 2]$.

This particular function realization produces components $[\ell]$ containing between 2 and 8 components of $x$ in the $q = 30$ terms in (3). While the complexity of the function might have some effect on the computational time to carry out the MCMC, the timing is dominated by the size of the dataset $(x, y)$.

Table 3: Time to complete 20K MCMC iterations for a 200,000×40 dataset. The number of processors includes the master processor. The run time is wall clock time in seconds.

| processors | run time (s) | processors | run time (s) |
|---|---|---|---|
| 2 | 347087 | 24 | 9660 |
| 4 | 123802 | 30 | 6303 |
| 8 | 37656 | 40 | 4985 |
| 16 | 16502 | 48 | 4477 |

Table 3 shows the time required to carry out the MCMC draws as a function of the number of processors, using this parallel implementation of BART. Here a total of 20,000 MCMC iterations were carried out for each timing run. As expected, the running time decreases with the number of processors, and the speed-up is nearly linear – the run time is about half when the number of processors is doubled, for example when moving from 24 to 48 cores.

Figure 2 shows how the inverse of run time increases as a function of the number of processors. The vertical axis is proportional to the number of MCMC steps per unit time (hour). As such, if the algorithm scales linearly, we would expect the slope of this line to be constant as we increase the number of cores. This is largely what we observe in this figure, with a slight change occurring around 30 cores which may be attributable to the particular design of the computer used in this example.

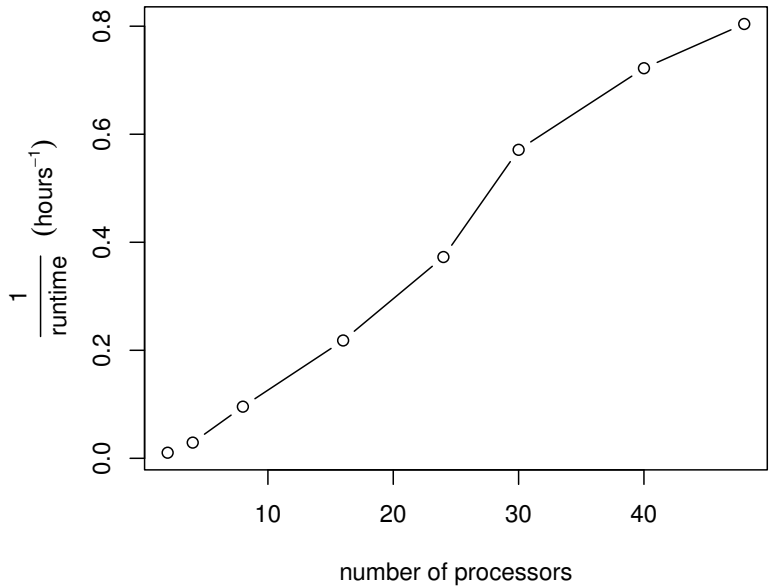

Figure 2: The inverse of wall clock time (in hours$^{-1}$) required to carry out 20K iterations of the parallel MCMC implementation for sampling the BART posterior distribution as a function of the number of processors. Here the dataset is $200,000 \times 40$, produced by Friedman's random function generator.

## 5 Scalability

We analyze the scalability of the proposed MCMC algorithm in terms of the notions of speedup and isoefficiency (e.g. Yero and Henriques (2007)). First, some basic quantities need to be defined. The speedup of an algorithm,

$$S(n, p+1) = \frac{T_{seq}}{T_{par}},$$

is the ratio of the times taken to run two instances of the algorithm. Typically, the speedup is regarded as the ratio of the sequential (or serial) algorithm's time to the parallel algorithm's time with $p+1$ cores, as expressed above. Alternatively, the speedup could instead be measured relative to a smaller number of parallel cores. Here, $n$ signifies the size of the problem, in our case the size of the dataset used in fitting the BART model. The efficiency,

$$E(n, p+1) = \frac{S(n, p+1)}{p+1},$$

is simply the speedup normalized to the number of cores used in the parallel version of the algorithm. For instance, if the algorithm were embarrassingly parallel (i.e. no communication overhead) then the parallel time would be given by $\frac{T_{seq}}{p+1}$, resulting in an efficiency of 1.0.

Table 4: Complexity of Serial and Parallel BART algorithms. The variable $p$ denotes the number of slave processor cores, $v$ the number of covariates, $n$ the total dataset size, $d$ is a random variable representing tree depth, $b$ is a random variable representing the number of bottom nodes in a tree and $\tau$ is a random variable representing the size of a tree. The approximate order per MCMC iteration is arrived at by assuming computational time for tree operations can be treated as constant, which we have found to be reasonable for BART due to the small tree sizes.

|  | Serial | Master Computation | Master Communication | Slave Computation | Slave Communication |
|---|---|---|---|---|---|
| Birth/Death | $O(n+nd+vd+d+\tau+bvd)$ | $O(\tau+d+vd+bvd)$ | $O(p)$ | $O(\frac{n}{p}+\frac{n}{p}d)$ | |
| Draw $\mu$ | $O(\tau+b+n+nd)$ | $O(\tau+b+pb)$ | $O(p+pb)$ | $O(\tau+b+\frac{n}{p}+\frac{n}{p}d)$ | $O(b)$ |
| Draw $\sigma$ | $O(n)$ | $O(p)$ | $O(p)$ | $O(\frac{n}{p})$ | |
| per MCMC | $O(mn+mnd+mvd+md+m\tau+mbvd+mb)$ | $O(m\tau+md+mvd+mbvd+mb+mpb+mp)$ | $O(mp+mpb)$ | $O(m\tau+mb+m\frac{n}{p}+m\frac{n}{p}d)$ | $O(mb)$ |
| approx. | $O(mn+mb)$ | $O(mp+mb+mpb)$ | $O(mp+mpb)$ | $O(m\frac{n}{p}+mb)$ | $O(mb)$ |

The isoefficiency function is defined as

$$I(p+1, e) = n_e,$$

where $e$ is the desired efficiency level with $p+1$ cores and $n_e$ represents the problem size to reach an efficiency $e$ with $p+1$ cores. This implicit function essentially relates the level of efficiency desired with the problem size, $n_e$, required to achieve that level of efficiency. Yero and Henriques (2007) define an algorithm to be e-Isoefficient if the efficiency of the algorithm with $p+1$ processors can be maintained at the level $e$ by increasing the problem size to some finite size $n_e$. This is relevant for algorithms involving Big Data as scaling to a large number of cores for a fixed problem size may be less practically relevant than scaling to a large number of cores for increasingly large datasets.

To determine if the BART MCMC algorithm is scalable in the sense of e-Isoefficiency, the speedup of the algorithm must be determined. This can be done entirely empirically, or by constructing a model for the speedup motivated by the theoretical complexity of the algorithm in question. In the later case, the runtime of an algorithm can be expressed as

$$\text{runtime} = \#\text{ of operations} \times \text{time per operation}$$

where the number of operations can be approximated by the order of the algorithm and the time per operation can be thought of as a machine-specific constant that maps the algorithmic order to the algorithms runtime.

As a simple example, consider the draw of $\sigma$ in our sampler which requires the residual sum of squares. In a serial algorithm, the order of this calculation is $O(n)$ and we can think of the runtime being $c_0 \times n$ for some constant $c_0$. In contrast, the parallel algorithm consists of the slave codes each calculating the partial residual sum of squares, a calculation of order $O\left(\frac{n}{p}\right)$ which, we assume, happens simultaneously on all slaves. Subsequently, the results from the $p$ slaves are collected and added on the master node, a communication calculation of order $O(p)$. The runtime of each of these components of the parallel algorithm can be thought of as $c_1 \times \frac{n}{p}$ and $c_2 \times p$, where the constant $c_2 >> c_1$ since it involves communication operations which are slow relative to computational operations. An approximation of the speedup can then be calculated as the fraction of these serial and parallel runtimes.

In order to apply this concept to the entire BART MCMC sampler, the order of the serial and parallel algorithms are summarized in Table 4. In this table, we describe the algorithmic order for each of the MCMC steps, and then the overall algorithmic order for a single iteration of the MCMC. The complexity described relates to our particular implementation, however it may be that alternative implementations could have somewhat different theoretical complexity. For instance, many of the terms not involving $n, m$ or $p$ relate to tree operations, which may be absent in other implementations which sacrifice greater memory usage in exchange for less computational overhead. Since we find that most tree operations are fast due to the shallow depth of BART's trees, we do not make this tradeoff in our implementation. However, in order to construct a model for speedup motivated by the complexity, it makes sense to treat tree operations as constant in order to identify the main factors that affect the speedup, which we have done in the "approx" row in Table 4.

Motivated by the approximate complexity of the BART MCMC sampler, a reasonable starting model for the speedup of BART can be found by forming a linear model with interactions for the serial runtime (involving regressors $n, m$ and $b$) and similarly for parallel runtime (involving regressors $\frac{n}{p}, m, p$ and $b$):

$$S(n, p+1) = \frac{T_{seq}(n, m, b)}{T_{par}(\tilde{n}, m, b, p)}, \tag{4}$$

$$T_{seq}(n, m, b) = \alpha_1 m + \alpha_2 n + \alpha_3 mn + \alpha_4 mb + \alpha_5 nmb,$$

$$T_{par}(\tilde{n}, m, b, p) = \beta_1 m + \beta_2 \tilde{n} + \beta_3 p + \beta_4 b + \beta_5 m\tilde{n} + \beta_6 mp + \beta_7 mb$$

$$+ \beta_8 \tilde{n} p + \beta_9 \tilde{n} b + \beta_{10} pb + \beta_{11} m\tilde{n} b + \beta_{12} mpb$$

$$+ \beta_{13} m\tilde{n} p + \beta_{14} m\tilde{n} pb,$$

where $\tilde{n} = \frac{n}{p}$ is taken as the problem size on each slave node in the parallel implementation of the MCMC algorithm, and the other variables are as defined previously. Here, the $\alpha$'s and $\beta$'s are the unknown machine-specific parameters.

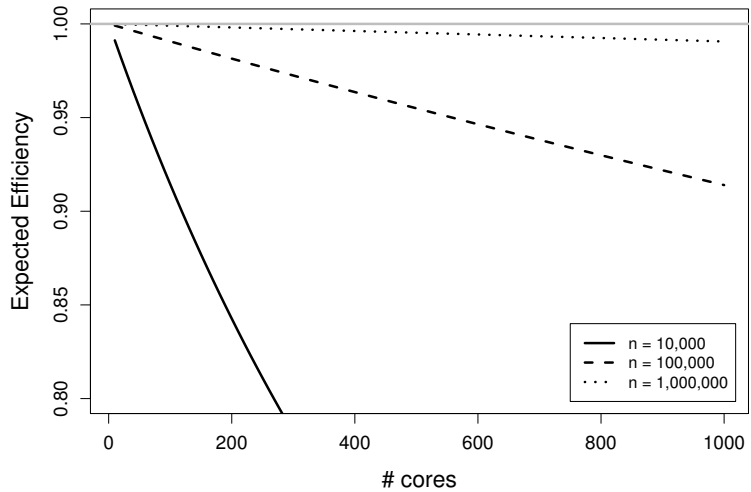

Figure 3: Expected efficiency of the BART MCMC algorithm for three problem sizes. The horizontal solid grey line represents the maximum efficiency of 1.0.

A simplified prior analysis considers the terms in the approximate algorithmic order of Table 4 by setting the corresponding coefficients to 1 in (4) while setting the remaining coefficients to 0. This is akin to considering speedup in terms of algorithmic order rather than having a dependence on the actual machine(s) used. A unique characteristic of the proposed MCMC algorithm is that the speedup depends on the random variable $b$. Determining scalability in such a situation does not appear to have been explored in the literature. We consider the notion of Expected e-Isoefficiency by determining the e-Isoefficiency when utilizing the expected speedup,

$$\mathsf{E}\left[E(n,p+1)\right] = \frac{1}{p+1}\mathsf{E}\left[S(n,p+1)\right] \qquad (5)$$
$$= \frac{1}{p+1}\int_b \frac{T_{seq}(n,m,b)}{T_{par}(\tilde{n},m,b,p)}\pi(b),$$

where the expectation is taken with respect to the prior distribution on the number of terminal nodes, $\pi(b)$. Although this distribution is not known in closed form, samples from it can be easily constructed by drawing from the prior distribution of node depth, $\pi(d)$.

The resulting expected efficiency curves for three problem sizes are shown in Figure 3. While the particular scaling of these curves from this prior analysis is not practically relevant, the plot does indicate that the proposed algorithm is efficient and scalable. This is seen by the expected efficiency curves with an increased problem size always lying above efficiency curves with smaller problem sizes. That is, if we increase the number of cores, we can maintain a desired level of efficiency by increasing the problem size accordingly.

A more practically relevant exercise is to perform a similar analysis using the equations in (4) on timing data obtained from a real machine while fitting a real dataset. We performed such an experiment on a 32 core computer running Linux using the Friedman function generator with the full factorial of experimental settings shown in Table 5. BART was fit to each run using 1000 MCMC iterations with the first 500 discarded as burn-in. In addition to serial runs, we ran the parallel MCMC sampler using 9, 17 and 25 cores at each of the experimental settings listed, and then fit linear models to the serial and parallel times. The linear models were arrived at by performing backward elimination starting from the full model (4) and using RMSE as the criterion. In this case, instead of averaging over the prior for $b$, we plug-in the average number of bottom nodes $\bar{b} = \sum_{i=1}^{N}\sum_{j=1}^{m} b_{ij}$ from the $N$ posterior samples for each fitted model.

Table 5: Experimental settings for scalability experiment using the Friedman function generator.

| Variable | Settings | | |
|---|---|---|---|
| m | 50 | 100 | 200 |
| n | 100,000 | 500,000 | 1,000,000 |
| p+1 | 9 | 17 | 25 |

The resulting linear models, shown in Figure 4 for serial and parallel times exhibit excellent fit to the observed runtimes of the algorithms, having an $R^2 > 0.9$. This indicates that the theoretical approximate complexity of Table 5 identifies the important components explaining the behaviour of our algorithms. At the same time, the linear models arrived at by backward elimination did not exactly match those used in our simplified prior analysis, indicating that the practical scaling of the algorithm using real data on a real machine is more complex. Nonetheless, the efficiencies from our experiment shown in Figure 5 do indicate that as the size of the problem increases we can increase the scalability of the parallel MCMC algorithm. One would expect this behaviour to eventually find a practical limit as machine limitations eventually become dominant.

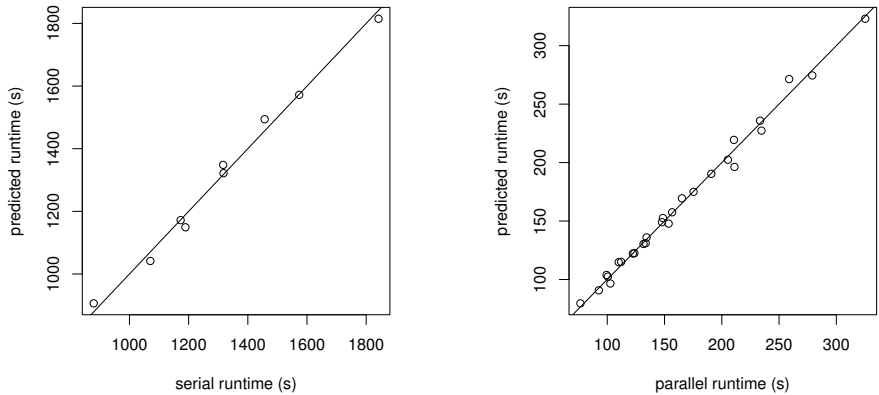

Figure 4: Fit of linear models to serial and parallel runtimes from the Experiment of Table 5. The fitted linear model for serial runtime was $T_{ser} = -2.046 \times m + 2.037e01 \times b + 1.282e-4 \times mn$ while the model for parallel runtime was $T_{par} = 2.011 \times b + 1.254e-4 \times m\tilde{n}$.

## 6 Prediction and Sensitivity Analysis

Predictions of the function $f$ at unobserved input settings $x^*$ can be constructed using the sampled posterior. For example, the posterior mean for $f$ can be estimated by

$$\hat{f}(x^*) = \frac{1}{N} \sum_{i=1}^{N} \sum_{j=1}^{m} g_j(x^* | T_j^i, M_j^i)$$

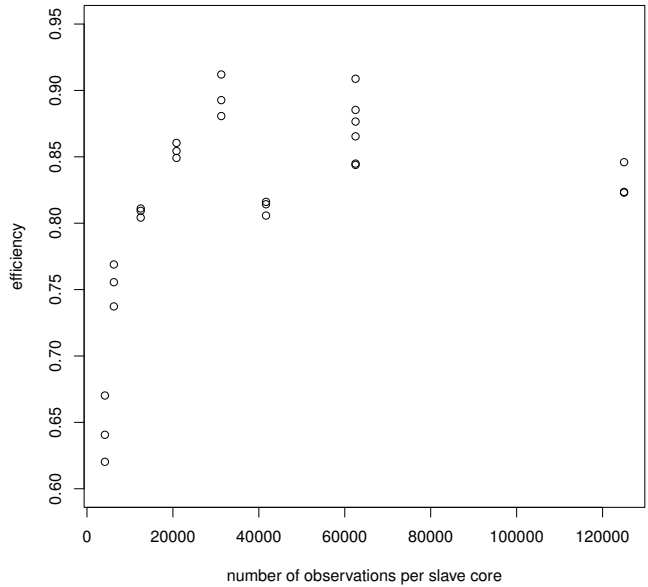

Figure 5: Parallel MCMC efficiency as a function of the number of observations per slave core, $\tilde{n}$, for the experiment of Table 5.

where $i$ indexes the $N$ MCMC draws. Uncertainty bounds can be similarly calculated. Such computations are embarrassingly parallel by simply subsetting the inputs $x^* = (x^*_{(1)}, \ldots, x^*_{(p)})$ over $p$ computational cores and performing the predictions (or other calculations) for these subsets independently on each core.

Calculation of main effect functions (Cox, 1982; Sobol', 1993) or sensitivity indices (Saltelli et al., 2008) can also be performed efficiently using the predicted response with some minor communication overhead. Sobol's functional ANOVA decomposition uniquely represents $f(x)$ as the sum of increasingly complex terms

$$f(x) = f_0 + \sum_{k=1}^{d} f_k(x_k) + \sum_{1 \leq k < \ell \leq d} f_{k\ell}(x_k, x_\ell) + \cdots + f_{1 \cdots d}(x_1, \ldots, x_d).$$

The functions are computed by integrals over the $x$-space, so that

$$f_0 = \int_{[-1,1]^d} f(x)dx \text{ and } f_k(x_k) = \int_{[-1,1]^{d-1}} f(x)dx_{-k} - f_0,$$

where $dx_{-k}$ includes all components of $x$ but the $k^{\text{th}}$. The above integrals can easily be approximated via Monte Carlo integration, drawing $x$'s uniformly over their domain, and using the posterior mean estimate

$\hat{f}(x)$.

Similarly, the 1-way sensitivity index $S_k$ for input $k$ is

$$\begin{aligned} S_k &= \frac{V_k}{V} \\ &= \frac{Var_{x_k}\left(E_{x_{-k}}(f|x_k)\right)}{Var(f)} \\ &\approx \frac{\int_{x_k} \hat{f}_k^2 dx_k}{\int_x \hat{f}^2(x)dx - \hat{f}_0^2}. \end{aligned}$$

Since the calculation of such indices involve integrals over the input space, there is some communication cost, but it is easily managed. Saltelli et al. (2008) approximate these calculations using Monte Carlo. For instance, the numerator can be calculated as

$$V_k \approx \sum_{j=1}^{n_s} \hat{f}(x_{j1}^a, x_{j2}^a, \ldots, x_{jd}^a) \times \hat{f}(x_{j1}^b, x_{j2}^b, \ldots, x_{jk}^a, \ldots, x_{jd}^b) - \hat{f}_0^2$$

using samples $a, b$ each of size $n_s$ from the input space. These samples hold a common, independent value for $x_{jk}$, but are otherwise independent. This calculation can be implemented in parallel by generating matrices $A_{(i)}$ and $B_{(i)}$ on the $i = 0, \ldots, p$ slave nodes where each row of a given matrix represents a randomly sampled point in the $d-$dimensional input space. Each matrix has approximately $\frac{n_s}{p}$ rows generated independently on each core. The samples can be drawn from a uniform distribution or a quasi Monte Carlo strategy may be used, such as a Sobol sequence. One must ensure that the matrices are unique on each core, so for a uniform sampling strategy the random number generator seed must be different on all the cores. The integrals can then be approximated using the above summation by computing partial sums with the generated samples on each core and communicating these partial sums back to the master node. The master node then averages the partial results to arrive at the final Monte Carlo approximation of the sensitivity indices. This same parallel procedure can be used to approximate the total sensitivity index, $S_k^T$, of which the exact required calculations are described in detail in Saltelli et al. (2008).

# 7 Examples

## 7.1 Friedman Function Generator

To demonstrate the usefulness of the parallel MCMC sampler, prediction, and sensitivity analysis algorithms described, we are motivated by problems in computer experiments where the datasets are increasingly large and high-dimensional and there is a lack of statistical methodology and software available to handle them. In order to evaluate a new method for modeling data from computer experiments, one often makes use of test functions such as the Goldprice function (Ranjan et al., 2008). Since these test functions are typically low-dimensional, we use a single realization from the Friedman function described earlier as an appropriate choice for evaluating performance in the high dimensional settings we are interested in.

Taking the 200,000×40 dataset described in Section 4, the MCMC was carried out using our parallel implementation of BART. Scatterplots of $y$ vs. each $x_k$ are given in Figure 6 for the first 10,000 rows of $(x, y)$. These scatterplots indicate that although the problem is high-dimensional, only a few inputs have a strong effect on the response, particularly $x_8$ and $x_{10}$.

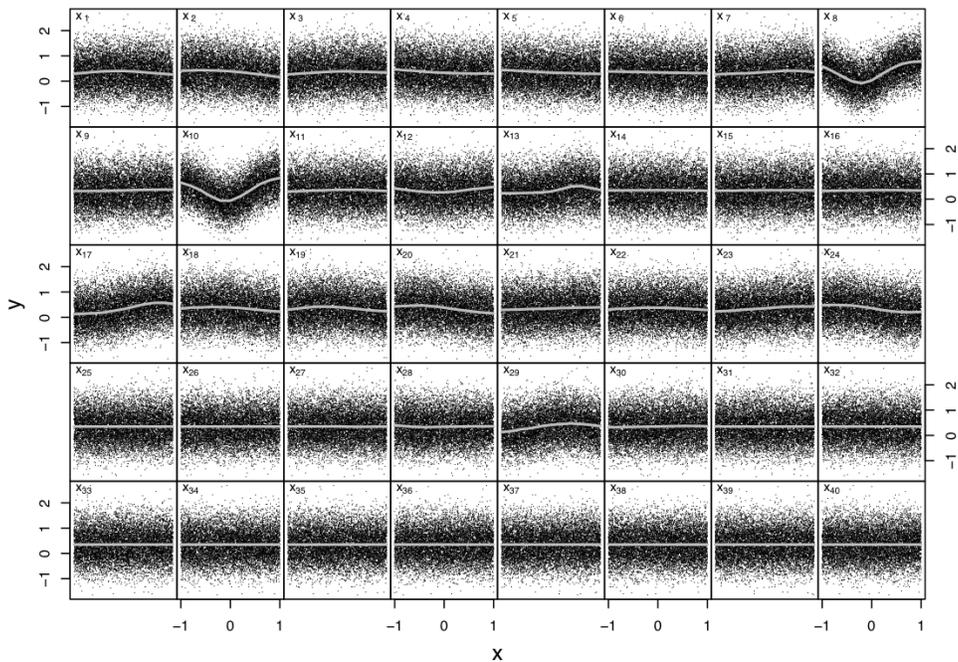

Figure 6: Scatterplots of each column of $x$ and the function output $y$. Of the 200,000 data realizations in the dataset, only the first 10,000 are shown here. The light lines give the main effect functions estimated from the fitted BART surface.

The MCMC was carried out on 48 processors for 500,000 iterations, with the first 100,000 being discarded for burn-in. The MCMC computation took a little over a day to complete, and would not have been possible to do using popular approaches in computer experiments such as Gaussian Process regression (Sacks et al., 1989), and while recent methods for large data in computer experiments have appeared (Haaland and Qian, 2011; Sang and Huang, 2012; Gramacy et al., 2013), they do not leverage parallel computing. From these 400,000 post burn-in realizations of the posterior, an equally spaced sample of 400 BART surfaces, each consisting of a sum of 200 trees, were saved to a file. These 400 posterior surfaces were then used to estimate main effect functions from a sensitivity analysis of the posterior mean surface and to predict a holdout set of $x's$.

The light lines in Figure 6 show estimates of the mean shifted main effect functions $\hat{f}_0 + \hat{f}_k(x_k)$, $k = 1, \ldots, 40$, as described in Section 6. Note that the sensitivity analysis has correctly identified $x_8$ and $x_{10}$ as being the most active inputs in this simulated example. The holdout predictions, calculated for a randomly drawn collection of 10,000 $x^*$'s, had an RMSE of 0.082, indicating a good fit.

## 7.2 WorldClim Minimum Temperature Dataset

The WorldClim dataset is a collection of interpolated temperature and precipitation fields on global land areas, excluding Antarctica (Hijmans et al., 2005). The data are freely available online at a variety of resolutions, resulting in very large datasets. We consider the 12 month minimum temperature field at a resolution of 10 minutes, which results in a dataset with 7,016,430 observations and a covariate space consisting of latitude, longitude and month (indexed as integers).

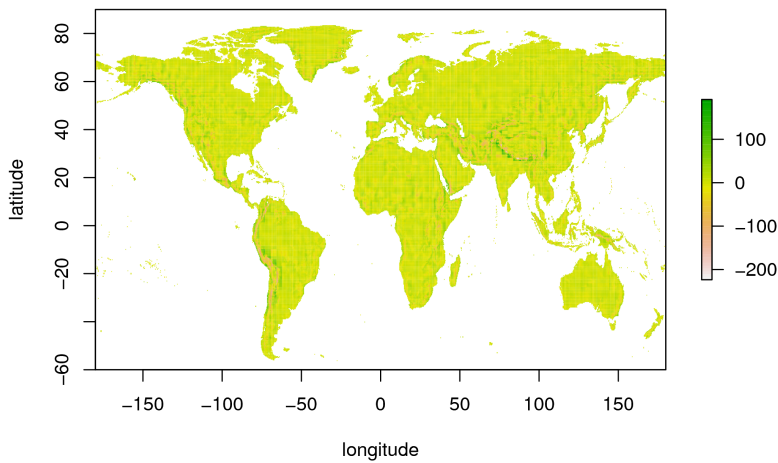

Figure 7: Residual of BART posterior mean for January WorldClim minimum temperature field in 10*degrees C at a resolution of 10 minutes.

Our aim here is not to illustrate a careful modeling of this dataset using BART, but to simply demon-strate the parallel MCMC algorithm's scalability on such a dataset. With this in mind, we ran the MCMC for 1,000 iterations, discarding the first 500 as burn-in. The algorithm was run on a public commodity cluster running Linux consisting of 48-core compute nodes, and the cluster was near full utilization (including other competing jobs) when we ran the test. Since the data is too large to obtain the runtime of the serial sampler, we ran the parallel sampler using 24, 48, 96 and 192 cores and compared the scalability of the MCMC with reference to the 24-core run.

Table 6: Efficiency of the parallel MCMC sampler with respect to a 24-core run for the WorldClim dataset.

| Number of Cores | 48 | 96 | 192 |
|---|---|---|---|
| Efficiency | 0.88 | 0.87 | 0.68 |

The resulting efficiencies observed for our runs are given in Table 6. They indicate high efficiencies of nearly 90% up to 96 cores. This means that the algorithm was achieving very good scalability up to 96 cores. For the 192-core run, we see the efficiency has dropped to just under 70%, reflecting the practical limits of our algorithm when run on the busy cluster with this particular dataset. Nonetheless, the performance is good even with this large number of cores, and had the cluster not been busy simultaneously serving other jobs, the efficiency would likely have been higher with the 192-core run.

While goodness of fit was not the aim of this exercise (especially with the small number of posterior draws), the residual field shown in Figure 7 demonstrates a reasonable fit to the data. The errors at extreme temperature values and the pixelation evident in the residual field suggests that a higher resolution of cutpoints is warranted for this dataset were a careful analysis undertaken.

## 7.3 Hockey Penalty Data

Abrevaya and McCulloch (2013) (henceforth AM) collected data on every penalty called in National Hockey League games from the 1995-1996 season to the 2001-2002 season. In ice hockey, the penalized team has to play with one less player for two minutes of a 60 minute game. AM's goal was to see if information in the game situation could be used to predict which team would get the next penalty. Their basic hypothesis was if the last few calls in a game were on the same team then it was highly likely that the next call would "reverse" and be on the other team. In addition, AM estimated the effect of several other factors such as the number of referees calling the game, whether the last penalty resulted in a goal, and which team is the home team. A variety of measures where also collected to characterize the different teams.

The resulting data set has 57,883 observations on 27 variables with each observation corresponding to a penalty call. The dependent variable is binary, indicating whether the call is a "reverse call" (i.e. the team penalized is different from the team penalized on the previous call). AM compared BART to Random Forests and Boosting and found BART did well. Since $p$(revcall), the probability that the call reverses, is never close to zero or one, AM simply used $y = f(x) + \epsilon$ where $y$ is 0 (no reverse call) or 1 (reverse call) and $p$(revcall) $= p$(revcall $|\, x) = p(y = 1\,|\, x)$. Here "$x$" consists of 26 variables characterizing the game situation and the two teams playing the game.

With 60 thousand observations, the `bart`/`BayesTree` version of BART is unbearably slow (see Table 1). AM used the new serial version of BART discussed in Section 3.2. This version took 3.5 hours to run 25,000 iterations. While several runs could be done overnight in parallel, this run time discouraged AM from considering a variety of BART specifications, instead using the defaults. It is of interest to assess whether the results are sensitive to the BART prior/model defaults. In assessing the predictive performance of BART, CGM used cross-validation to choose the BART specification and found this outperformed the default choice.

Using a 32-core machine we ran 4 different prior/model specifications using 8 processors for each run. This gives 8,269 observations per slave. Figure 5 suggests this may be a reasonable allocation. And, using the fitted linear models of Figure 5 to approximate the expected efficiency (5) suggests a maximum efficiency around 0.74 when using 6-9 cores (with $m = 200$ and the overall average $\bar{b} = 18$ from these earlier experiments). Using this setup, the parallel version is 6.34 times faster giving an actual efficiency of 0.8. It takes about 30 minutes to run 25,000 iterations as opposed to 3.5 hours. A single run could be made faster by using more than 8 cores, but the efficiency will diminish as fewer observations are allocated to each slave and this way all 4 specifications could be run at the same time on the 32 core machine. In this particular instance, a 30 minute wait time was convenient for the investigator.

An important BART prior parameter is `kfac` which determines the amount of shrinkage on bottom node means. The default is `kfac=2`. We tried `kfac=1` and `kfac=3`. Another basic choice in using BART is the number of trees in the sum ($m$ is Section 2). The default is 200. We tried 100 and 500. So, our four runs correspond to `kfac` equal 1 or 3 and the number of trees equal 100 or 500. Other parameters are held at the defaults.

Figure 8 reproduces a figure from AM and adds in estimates from the new four new runs. The labels on the horizontal axis indicate a game scenario. `tworef = 0` or 1, indicates whether the game was called by one referee or two; `ppgoal` = 0 or 1, indicates whether a goal was scored on the last penalty; `home` = 0 or 1, indicates whether the last penalty was on the home team. `inrow2=0` means the last two calls were on different teams. `inrow2=1` means the last two calls were on the same team. `inrow3=1` means the last three calls were on the same team. `inrow4=1` means the last four calls were on the same team. The black dot and solid vertical line above each label indicate the posterior mean and 90% posterior interval of $p$(revcall) given the game scenario and the default specification. The posterior means from our four new runs are plotted using text symbols to indicate the specification. For all but the last two game scenarios, the posterior means from new runs are at almost the same level as the posterior mean from the default. In the last two scenarios, the `kfac=1` (`k1`) and number of trees equal 500 (`m500`) results are substantially higher. These two situations (three or four calls in a row on the same team) occur very infrequently, so we have the plausible result that in those situations where we have less data, the specification is influential. At `inrow4=1`, the posterior mean of $p$(revcall) is 0.75 using the default specification and 0.8, 0.73, 0.78, and 0.74 at `kfac=1`, `kfac=3`, `m=100`, and `m=500`, respectively. Fortunately for AM, the size of the effects of the `inrow` variables is so large that the sensitivity to the specification does not affect the basic conclusions.

## 8 Conclusion

We have presented and implemented a straight-forward SPMD approach for sampling the posterior distribution resulting from BART. In addition we have also constructed post-processing parallel code to carry out basic sensitivity analyses and prediction.

The novelty of the model implementation we have described is in using the ideas of sufficiency and data reduction inherent in a statistical model to create a parallel MCMC sampler that can efficiently sample the posterior distribution when dealing with large datasets. This sampler has a number of unique properties, such as the ability to work with observational datasets which may be too large to be stored in a single contiguous location.

To evaluate the scalability of the proposed sampler, we introduce the notion of Expected e-Isoefficiency, and determine through simulation under our default prior that the algorithm can scale to handle massive datasets on large parallel computers. We were then able to confirm this behaviour through an empirical experiment using simulated datasets with up to 1 million observations. This is all achieved within the usual Bayesian framework.

Motivated by the theoretical complexity of our algorithm, we showed through our empirical experiment how one can estimate the serial and parallel runtime of our samplers using linear models. This is useful in allocating parallel computation resources that maximize efficiency.

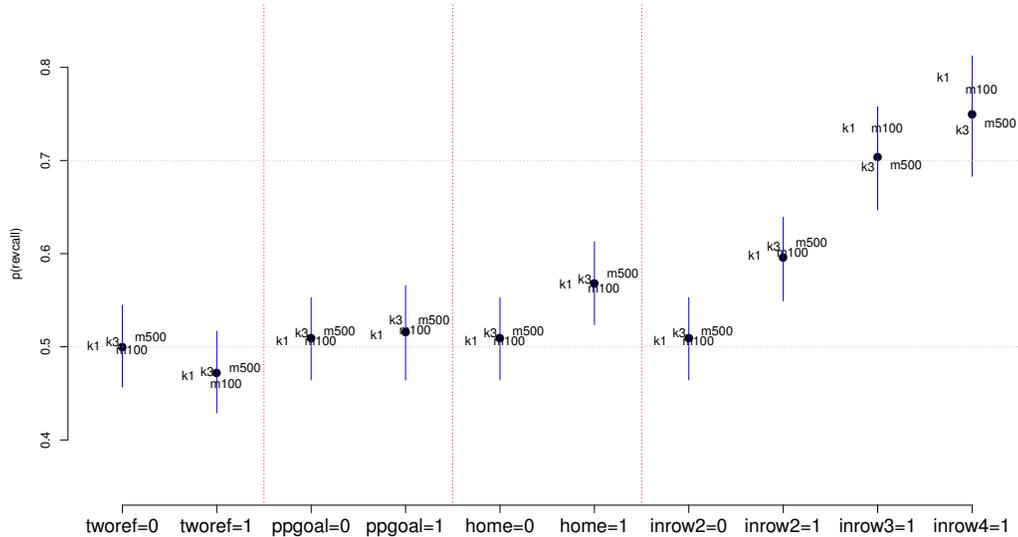

Figure 8: Posteriors of $p$(revcall) given 10 different game scenarios. The scenario is indicated by the label on the horizontal axis. The posterior mean obtained using the default specification is indicated by a solid black dot. The vertical line through dot indicates a 90% posterior interval using the default. Posterior means from our 4 new runs are plotted with the text symbols: `k1:kfac=1, k3:kfac=3, m100:100 trees, m500:500 trees`.

We also demonstrate the capabilities of the algorithm by applying it to a dataset generated by the Friedman function, and observed that the sampler scaled nearly linearly up to 48 processor cores in this example. A more challenging example analyzing a 7 million observation WorldClim dataset demonstrated the algorithms ability to scale up to 100 cores on a busy public cluster. Finally, for the moderately sized hockey penalty data, the ability to choose an efficient number of processors allowed effective use of resources in studying the impact of BART's prior specification.

The code, written in C++ using MPI, is available at `http://www.rob-mcculloch.org`.